\pdfoutput=1 % only if pdf/png/jpg images are used
\documentclass[cite]{JINST}

\usepackage{dcolumn}  % Align table columns on decimal point

\let\ifpdf\relax
\usepackage{multirow}
\title{Principal Component Analysis of Cavity Beam Position Monitor Signals}

\author{Y.~I.~Kim$^a$, 
S.~T.~Boogert$^b$, Y.~Honda$^c$,  A.~Lyapin$^b$\thanks{Corresponding author.},
H.~Park$^d$,
 N.~Terunuma$^c$, T.~Tauchi$^c$
and J.~Urakawa$^c$ \\
\llap{$^a$}John Adams Institute at University of Oxford,\\
Oxford, OX1 3RH, UK\\
\llap{$^b$}John Adams Institute at Royal Holloway, University of London,\\
Egham, Surrey TW20 OEX, UK\\
\llap{$^c$}High Energy Accelerator Research Organization (KEK),\\
1-1 Oho, Tsukuba, Ibaraki Prefecture 305-0801, Japan\\
\llap{$^d$}Kyungpook National University, Dept. of Physics,\\
1370 SanKyuk-dong, Buk-gu, Daegu 702-701,South Korea\\
E-mail: \email{Alexey.Lyapin@rhul.ac.uk}}

\abstract
{
Model-independent analysis (MIA) methods are generally useful for analysing complex systems in which relationships between the observables are non-trivial and noise is present. 
Principal Component Analysis (PCA) is one of MIA methods allowing to isolate components in the input data graded to their contribution to the variability of the data.
In this publication we show how the PCA can be applied to digitised signals obtained from a cavity beam position monitor (CBPM) system on the example of a 3-cavity test system installed at 
the Accelerator Test Facility 2 (ATF2) at KEK in Japan. We demonstrate that the PCA  based method can be 
used to extract beam position information, and matches conventional techniques in terms of performance, while requiring considerably less settings and data for calibration.
}


\begin{document}

\section{Introduction}\label{sec:intro}

The Accelerator Test Facility 2 (ATF2)~\cite{atfweb},~\cite{atf2proposalVol1},~\cite{atf2proposalVol2} at KEK has been designed to prove the principle of the compact final focus beam optics design based on the local chromaticity correction~\cite{finalFocus} required for future Linear Colliders (LCs) such as the International Linear Collider (ILC)~\cite{ilc} and Compact Linear Collider (CLIC)~\cite{clic}. LCs require 
precision transverse position measurements for a number of applications including, but not limited to, such crucial tasks as beam based alignment and beam optics tuning.  
Cavity beam position monitors (CBPM) have been proposed as the primary technology for 
the high resolution position measurements. 
Resolutions of a few hundred nanometre have been routinely demonstrated in installations of tens of such devices~\cite{kim} with 10 $\sim$ 50~nm 
achieved in smaller, limited range high gain systems~\cite{kim},~\cite{honda},~\cite{kimthesis}.
The ATF2 is a test facility of various beam diagnostics systems such beam position monitor (BPM) 
system, laser wire (LW) and optical transition radiation (OTR). All data used for this paper were taken at the ATF2.

\subsection{Cavity Beam Position Monitor}\label{sec:cavity}
When a charged particle beam passes through a cavity, various electromagnetic modes are excited. 
The excited electromagnetic fields are defined by the cavity shape and trajectory of the passing beam. 
Some of the excited modes are dependent on the transverse position of the beam. Hence, the beam position can be determined by selecting and measuring the strength of 
these modes. Usually, the first dipole mode is used for position measurements as it has the strongest beam coupling among the position dependent modes and also flips the phase of the oscillations by 180$^\circ$ when the offset changes its sign relative to the electric centre of the cavity, which can be detected using an external phase reference. 

\begin{figure*}[htb]
\includegraphics[width=150mm]{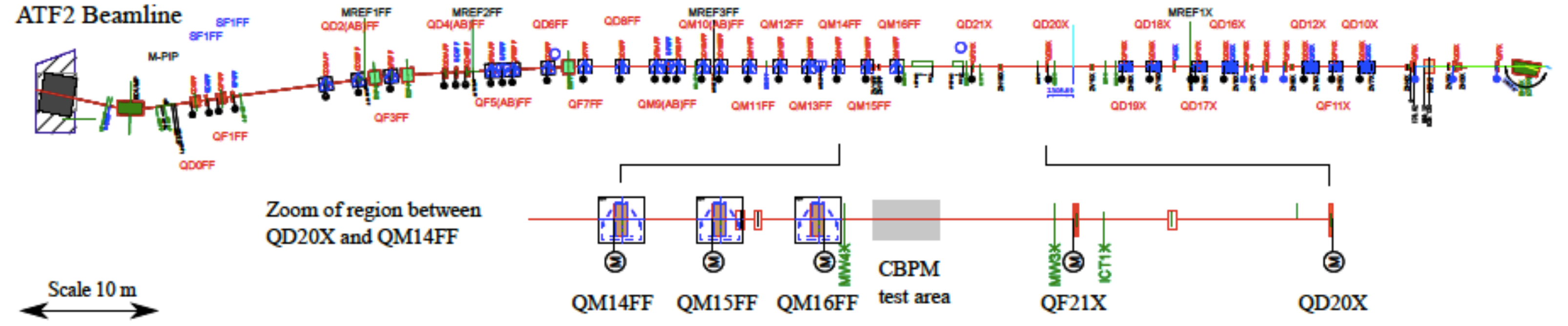}
\caption{\label{fig:figure1}The ATF2 beam line with a zoom in around the test location.}
\end{figure*}

\begin{figure*}\centering
\includegraphics[width=150mm]{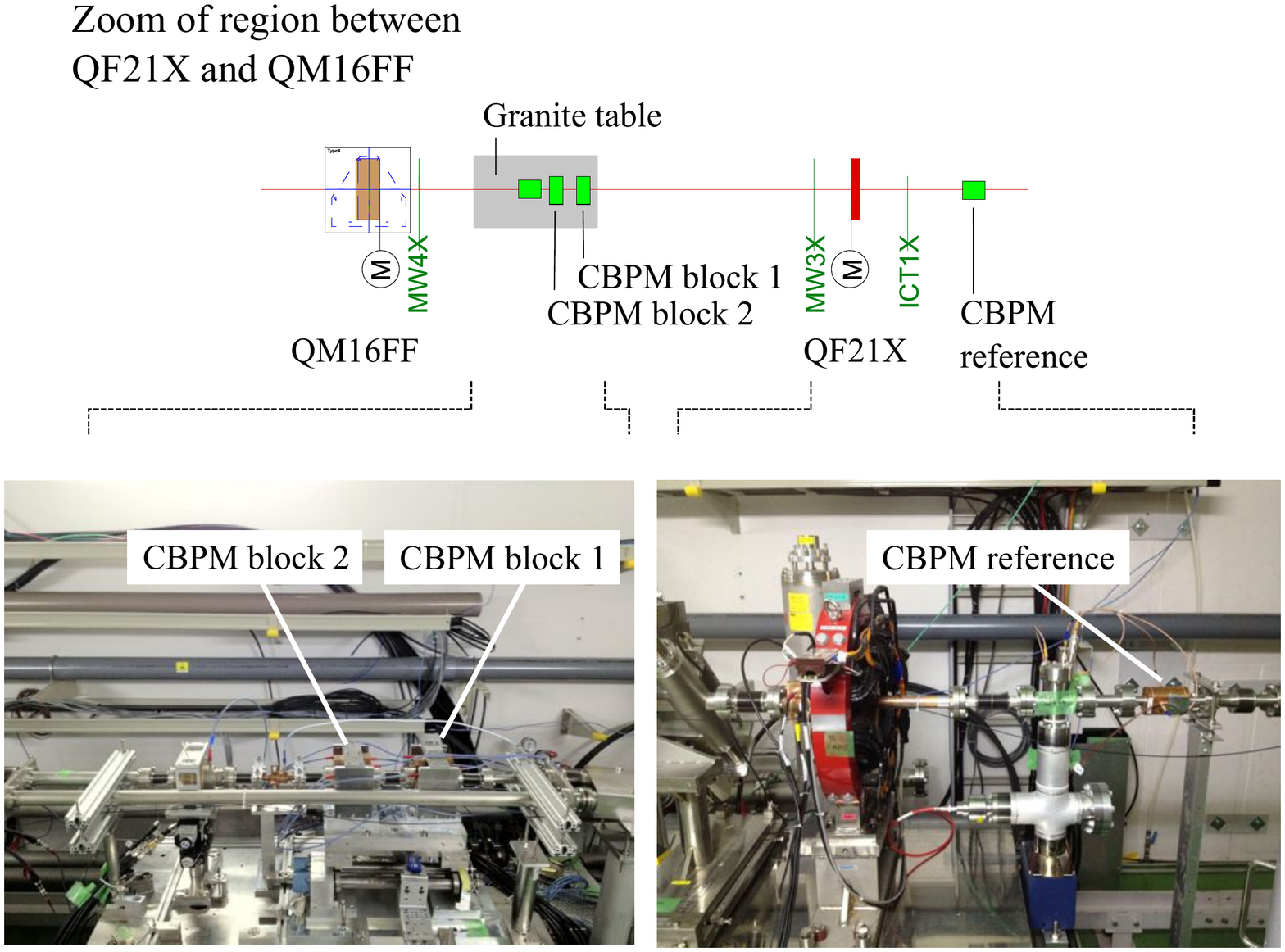}
\caption{\label{fig:figure2} Detailed schematic of the CBPM test region and photographs of the installed system.}
\end{figure*}

Figure~\ref{fig:figure1} shows the ATF2 beam line and Figure~\ref{fig:figure2} a schematic and photographs of the area 
where the test system was installed. The system located between quadrupoles QF21X and QM16FF of the ATF2 extraction beam line 
consisted of 2 blocks containing 2 CBPMs each, but only 3 of the 4 cavities were read out.  
Additional instrumentation included a tuned in frequency reference cavity.

The cavities constituting the test system used for analysis in this paper are rectangular in shape to 
split the $x$ and $y$ dipole modes and separate them in frequency thus reducing the cross-coupling between them. These cavities are coupled via rectangular 
slots into waveguides with coaxial adaptors. This arrangement allows 
the extraction of the position sensitive dipole mode and suppresses the strong 
monopole modes. 
The geometry is symmetric with a pair of couplers for each transverse plane.

A cavity outputs an exponentially decaying sine wave with angular frequency $\omega$
and decay constant $\tau$ defined by the geometry and material of the cavity. Assuming the bunch length is fixed, the dipole output 
voltage $V_{\rm d}(t)$ is given by 
\begin{equation}
\label{eq:dipole}
V_{\rm d}(t;x,\alpha,\theta) = \left[S_x x - j S_{\alpha}\alpha+ j S_{\theta}\theta \right] q e^{-t/2\tau_{\rm d}} e^{j(\omega_{\rm d} t+\phi_{\rm d})} ,
\end{equation}
where $S_x$, $S_\alpha$ and $S_\theta$ are the sensitivities to the beam position 
$x$, bunch tilt $\alpha$ and beam trajectory $\theta$ respectively, and $q$ is the bunch charge. 
The phase of the signal $\phi$ depends on the bunch arrival time, and so is arbitrary unless an additional phase reference is used. 
This is usually provided by a reference cavity, operating the monopole mode at the same frequency as the dipole cavities. It also serves as 
an independent measurement of the bunch charge required for the position determination. The reference output voltage is
%In addition, a reference cavity, operating with the monopole mode at the same frequency as the dipole cavities, 
%provides an independent measurement of the bunch charge, 
%as well as a phase reference, so that these can be used in the position determination. The reference output voltage is  
\begin{equation}
\label{eq:reference}
V_{\rm r}(t) = S_{q} q e^{-t/2\tau_r} e^{j(\omega_{r}t+\phi_r)}.
\end{equation}
Note that the difference between $\phi_d$ and $\phi_r$ is fixed for any 2 points along the waveforms even if the frequencies do not match precisely.

Table~\ref{tab:ipbpmParameter} shows the design parameters of the cavities used in the experiment: the resonant frequency of the
dipole mode $f_{\rm d}$, coupling strength $\beta$, loaded quality factor $Q_{\rm L}$, 
internal quality factor $Q_0$, external quality factor $Q_{\rm ext}$, normalised shunt impedance $(R/Q)_0$ and decay time $\tau$. 
\begin{table}[hbt]
\centering
\begin{tabular}{c|c|c|c|c|c|c|c}
\hline \hline
Parameters     &  $f_{\rm d}$ [GHz] & $\beta$ &  $Q_{\rm L}$ & $Q_0$ & $Q_{\rm ext}$ &  $(R/Q)_0$ @ 1mm offset   & $\tau$ [ns]  \\
\hline 
$x$ dipole       & 5.7086           & 1.63         &  2067    & 5424    &  3335                & 0.549                                      & 58                \\ \hline
$y$ dipole       & 6.4336           & 3.32        & 1217       & 5459   & 1586                & 1.598                                      & 30               \\
\hline \hline
\end{tabular}
\caption{Simulated parameters of the CBPMs~\cite{honda}.}
\label{tab:ipbpmParameter}
\end{table}

\subsection{Signal Processing and Calibration}\label{sec:signal}

\begin{figure}[htbp]
\begin{picture}(100.0, 190.5)
\put(100.0,190){\small a)}
\put(330.0,190){\small b)}
\put(-5.0,0)
{\includegraphics[width=7.5cm]{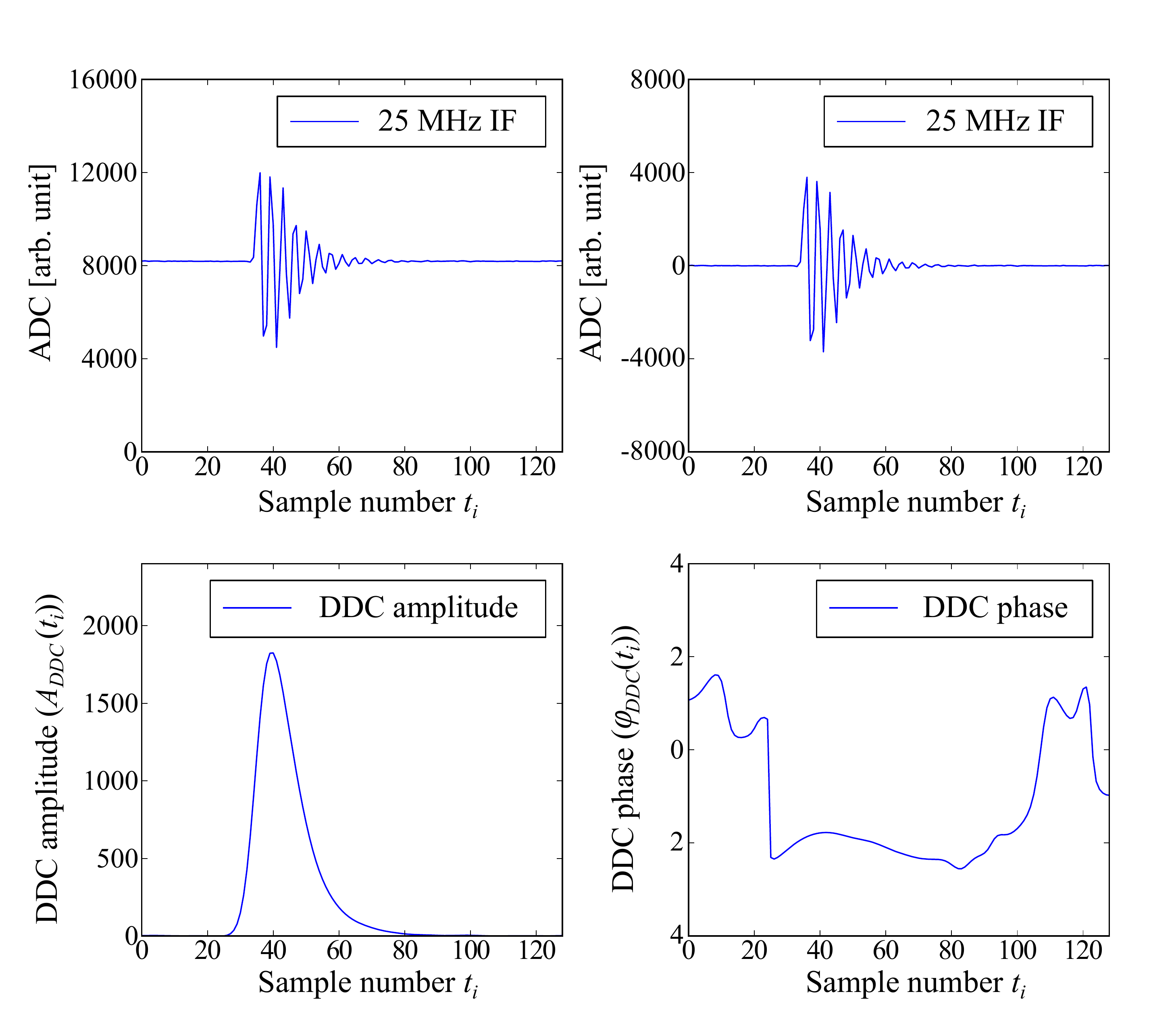}}
\put(210.0,0)
{\includegraphics[width=8.6cm]{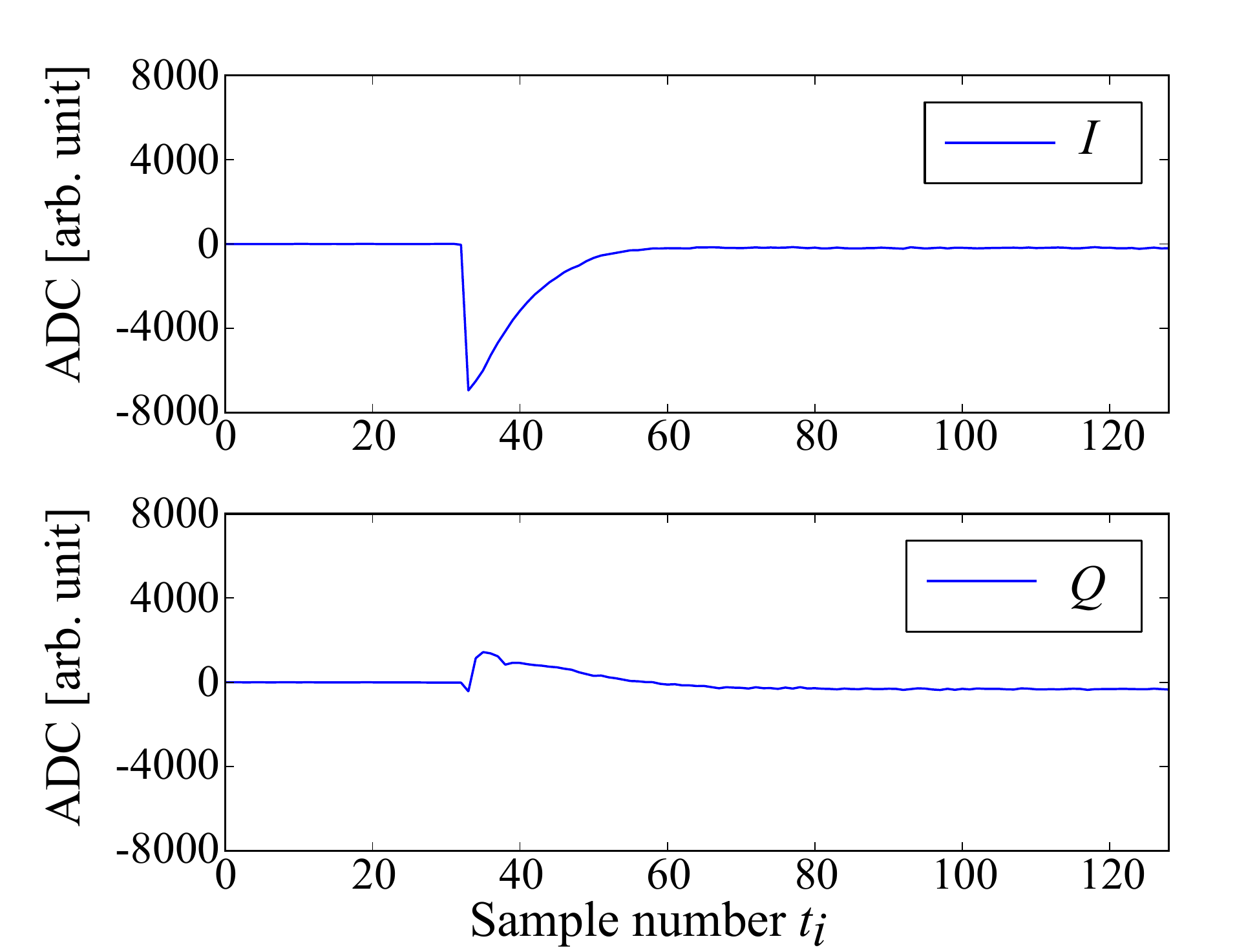}}
\end{picture}
\caption{Example of a) nonzero IF digitised and digitally down-converted signals for a dipole cavity
and b) zero IF digitised $I$ and $Q$ signals.}
\label{fig:figure3}
\end{figure}

The high frequency cavity output is most commonly down-converted to a more manageable intermediate frequency (IF), about tens of MHz , followed by an 
additional digital downconversion (DDC) stage, or directly to the "baseband" ("zero IF"). In any case, the target is to obtain the amplitude and phase envelope of the position signal normalised to the reference. 
We used both methods as the bandwidth of the processing electronics allowed us to do so. The down-converted signals were digitised at 100~MS/s by 14-bit digitisers. 
Figures~\ref{fig:figure3} show example waveforms processed in both zero and nonzero IF configurations. 
The top left plot in Figure~\ref{fig:figure3} (a) is a 25 MHz down-converted raw digitised signal.
The top right plot is the same signal with the pedestal subtracted. 
The signal is then mixed with a digital local oscillator (LO) and filtered to give the amplitude and phase of the signal shown 
in the bottom left and right plots respectively~\cite{kim}. The phase incursion along the waveform is minimised by adjusting the frequency of the digital LO to reduce the effects of the trigger jitter. The plots in Figure~\ref{fig:figure3} (b) show the real, "in-phase" ($I$), and imaginary, "quadrature" ($Q$) components of the extracted phasor. For position calculation, the phasor is sampled at a single time $t_{\rm s}$, roughly one filter length after the amplitude peak.
\begin{equation}
I_{\rm s}+jQ_{\rm s} = g\frac{V_{\rm d}(t=t_{\rm s,d})}{V_{\rm r}(t=t_{\rm s,r})} \, ,
\end{equation}
where the phasor $g$ accounts for any differences between the dipole and reference processing.

The phasor $I+jQ$ needs to be rotated by an angle $\theta_{IQ}$,  $I^{\prime}+jQ^{\prime} = e^{i\theta_{IQ}}(I+iQ)$, so that its in-phase component $I^{\prime}$ is proportional to the position and the quadrature component $Q^{\prime}$ only contains the angle and tilt information. The required rotation of the 
$IQ$ plane is measured during the calibration when a significant position variation is introduced to reduce the effect of the angular jitter.  
The position scale $S$ for converting $I^{\prime}$ into position is measured by offsetting the CBPMs by a known amount using the mover or an orbit bump.

\subsection{Principal Component Analysis}\label{sec:pca}

Methods of Model Independent Analysis (MIA) are used to extract relationships in noisy 
data without using an underlying model. In accelerator physics, MIA has been used, for example, to analyse 
the complex beam dynamics~\cite{mia} and extract beam position information~\cite{wang1}, and the number of applications is growing due to both improving availability of numerical algorithms and increasing complexity of the accelerator systems and their requirements. So, a method of MIA that 
does not require a specific machine model is important to make sensitive 
measurements for a future Linear Collider (LC) when the beam orbit changes with time~\cite{mia}.
So far, MIA have been applied to processed CBPM data to measure position 
resolution~\cite{kim},~\cite{walston} and improve calibration parameter determination~\cite{frankie}, but not to raw CBPM waveform data.

Principal component analysis (PCA) is a MIA that is used to reduce the dimensionality of the data.
The basic idea of PCA is to transform the raw data into a basis which best explains the variation within the data.
If the data is a matrix {\bf d} with $N$ variables in columns and $M$ rows of repeated measurement, 
it can be transformed using an orthogonal matrix $\bf W^{T}$ to ${\bf Y}$ given by 
\begin{equation}
{\bf Y} = \bf W^T {\bf d}.
\label{eq:pcaTransform}
\end{equation}
The matrix ${\bf W^T}$ can be 
considered as a rotation matrix that transforms the data into another linear vector space. 
The vectors ${\bf W^T}_{i*}$ form a set of $N$ basis vectors which the data is projected onto. 
The PCA method determines the transformation matrix ${\bf W^T}$ whilst keeping the variability of the original data.

PCA determines ${\bf W^T}$ in such a way to make the covariance matrix of ${\bf Y}$ a diagonal matrix. 
The covariance matrix of the transformed data ${\bf Y}$ is calculated by 
\begin{equation} \label{eq:pcaCov}
{\bf YY^T}  =  \left({\bf W^T d}\right) \left({\bf W^T} {\bf d}\right)^{\bf T} \\ \nonumber
                               =  {\bf W^T} {\bf dd^T} {\bf W}.
\end{equation}
The data matrix ${\bf d}$ can be decomposed using singular values decomposition (SVD),
\begin{equation}\label{eq:svd}
{\bf d} = {\bf U}{\bf S}{\bf V^T},
\end{equation}
where ${\bf S}$ is a diagonal matrix, ${\bf U}$ and ${\bf V^T}$ are orthogonal matrices of size $M \times M$ and $N \times N$, respectively. Using Equations~\ref{eq:pcaCov} and~\ref{eq:svd} the covariance matrix of ${\bf d}$ is 
\begin{equation} 
{\bf d d^T} = {\bf U}{\bf S^2}{\bf U^T} = {\bf W}{\bf YY^T}{\bf W^T}.
\end{equation}
So the PCA transformation matrix ${\bf W}$ can be identified with matrix ${\bf U}$ from the SVD of the data matrix ${\bf d}$, and the covariance matrix of the transform data is diagonal and can be identified with the singular value matrix squared. 

\section{Application of Principal Component Analysis to Cavity Beam Position Monitor data}\label{sec:pcabpm}

There are many software packages for calculating principal components, we used the Python implementation \texttt{scikit.learn}~\cite{scikit-learn}. 
PCA was applied to the calibration data, guaranteeing the largest variability in the waveform data and thus variance is the position dependent signal. The reference cavity waveforms have been processed in a similar manner to provide a beam charge measurement. 

\subsection{Principal Component Analysis of Cavity Beam Position Monitor waveforms}

\begin{figure}[tbp]
\centering
\includegraphics[width=100mm]{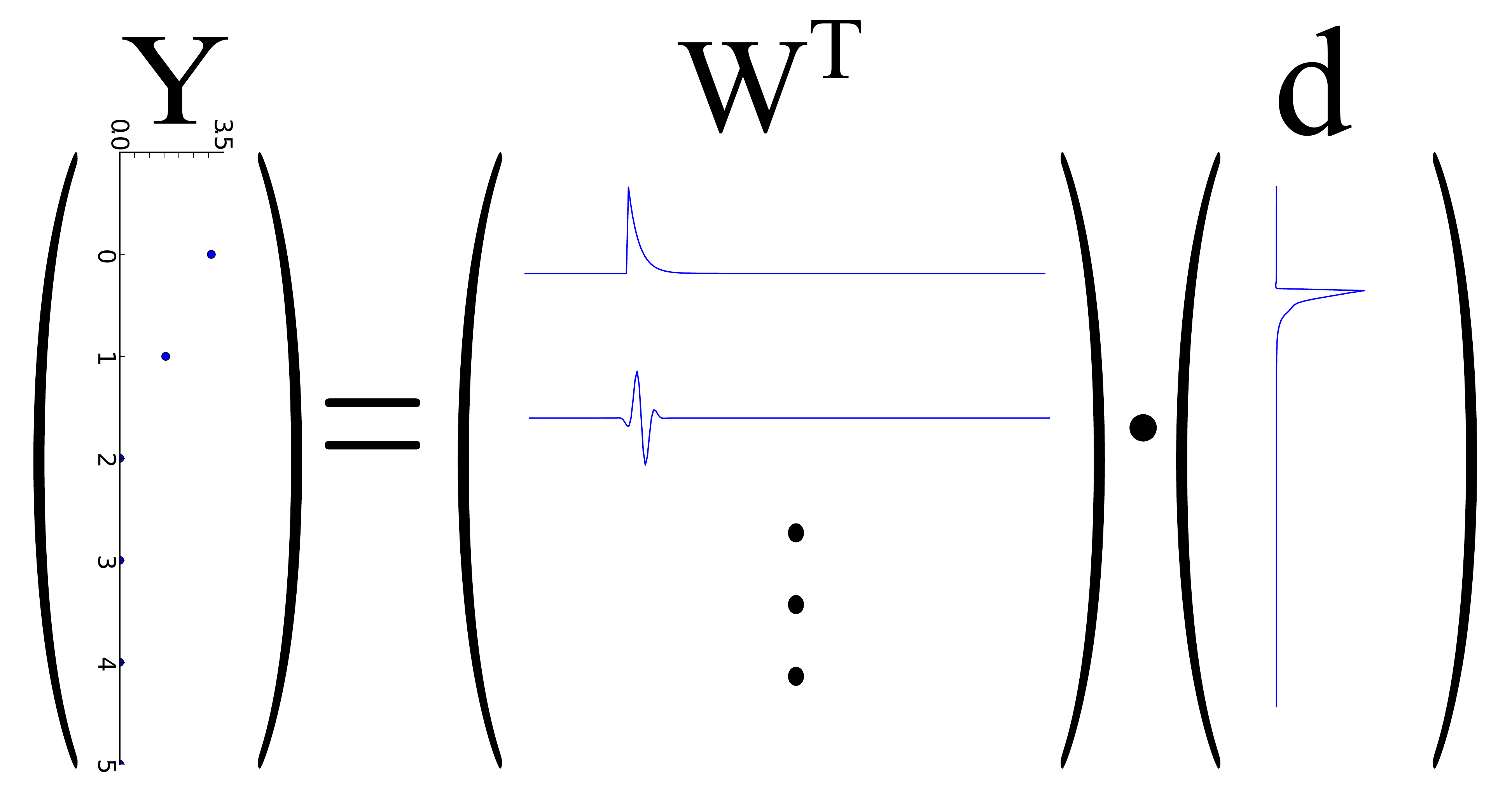}
\caption{An example of PCA on a zero IF signal.}
\label{fig:figure4}
\end{figure}

Data from a single CBPM for a single bunch machine pulse ${\bf d}$ is a vector of length $N$, where 
$N$ is the number of digitiser samples. 
The vector ${\bf d}$ may contain not only the wanted mode signal ${\bf d_{\rm d}}$, but also unwanted contributions 
from other modes and various noise sources ${\bf d_{\rm u}}$, so
\begin{equation} \label{eq:sigSum}
{\bf d} = {\bf d_{\rm d}} + {\bf d_{\rm u}}.
\end{equation}
In a measurement of the amplitude of ${\bf d}$, the unwanted signals contribute a systematic offset or noise. 
The vector ${\bf d}_{\rm d}$ varies depending on the beam position and charge, hence for calibration data it is expected to produce strong components in the PCA matrix, while noise sources will form higher components. 
This is shown diagrammatically in Figure~\ref{fig:figure4}, which is an example where the signal has zero IF. 

Unless $I\/Q$ demodulation is applied to the cavity signals, each cavity direction has one output.
Each sampled signal is then expressed as a linear combination of basis vectors
\begin{equation}\label{eq:linearSum}
d(t_i)  =  \sum_{j} y_j {\bf W}_{ji}^{\bf T},
\end{equation} 
\noindent where coefficient $y_j$ is the relative contribution of basis vector ${\bf W}_{j*}^{\bf T}$. 
They can be found by taking the dot product between $d(t_i)$ and
${\bf W}_{j*}^{\bf T}$. 

For CBPM data with nonzero IF, Equation~\ref{eq:linearSum} can be applied as follows:
\begin{equation}
V(t_i) =  \sum_{j} y_j \hat{V}_{ji}^{\bf T},
\end{equation}
where $\hat{V}_{j*}^T$ are the principal components of the data matrix of $V(t_i)$ measurements. 
For baseband-converted CBPM data, the analysis needs to be applied twice, once for the $I$ and once for $Q$ waveforms: 
\begin{eqnarray}
I(t_i) &  = & \sum_{j} y_j \hat{I}_{ji}^{\bf T}, \\
Q(t_i) & =  & \sum_{j} y_j \hat{Q}_{ji}^{\bf T}.
\end{eqnarray}

\begin{figure}[htbp]
\centering
\includegraphics*[width=100mm]{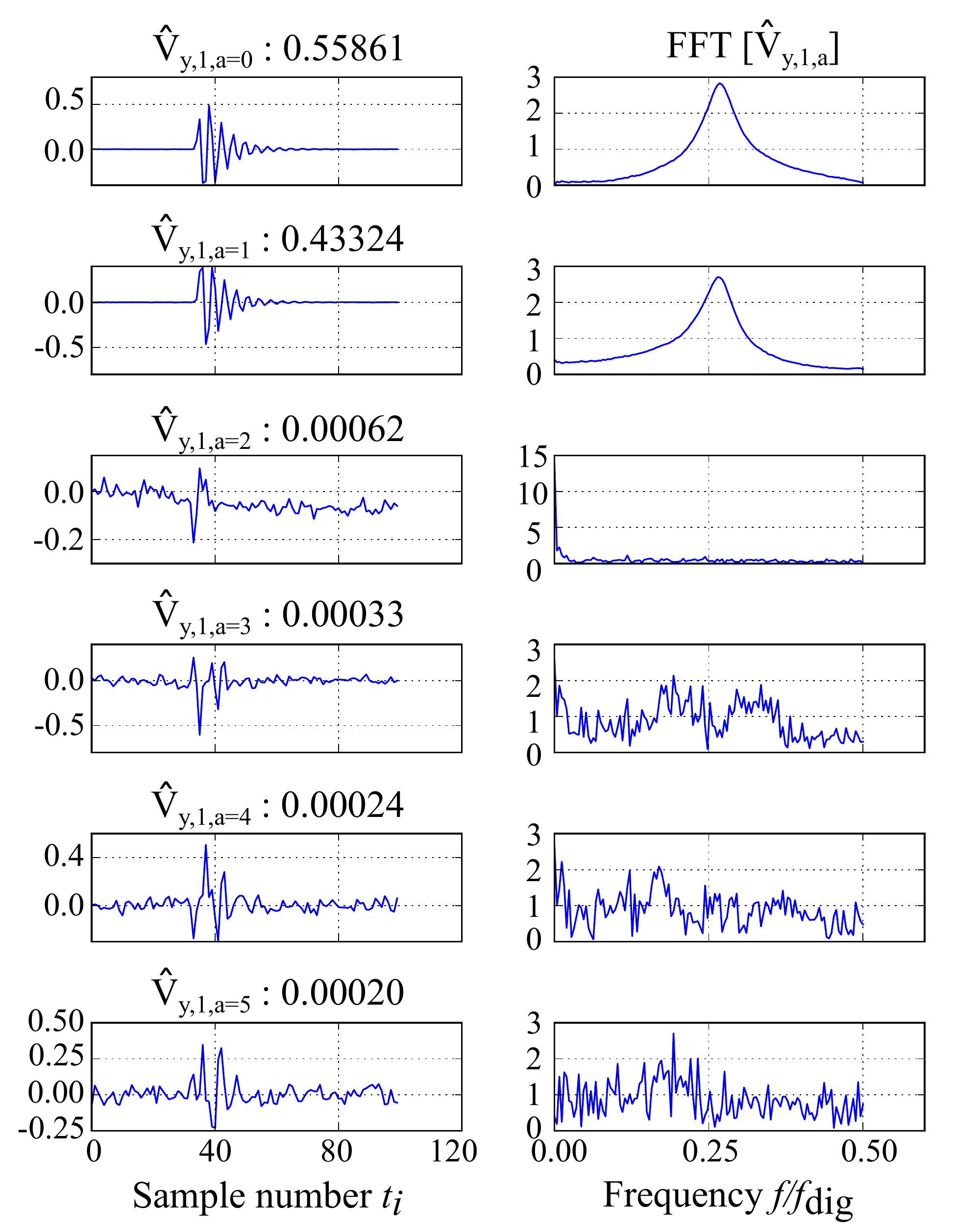}
\caption{PCA components for nonzero IF data (left column) and their FFTs (right column).}
\label{fig:figure5}
\end{figure}
\begin{figure}[htbp]
\centering
\includegraphics*[width=90mm]{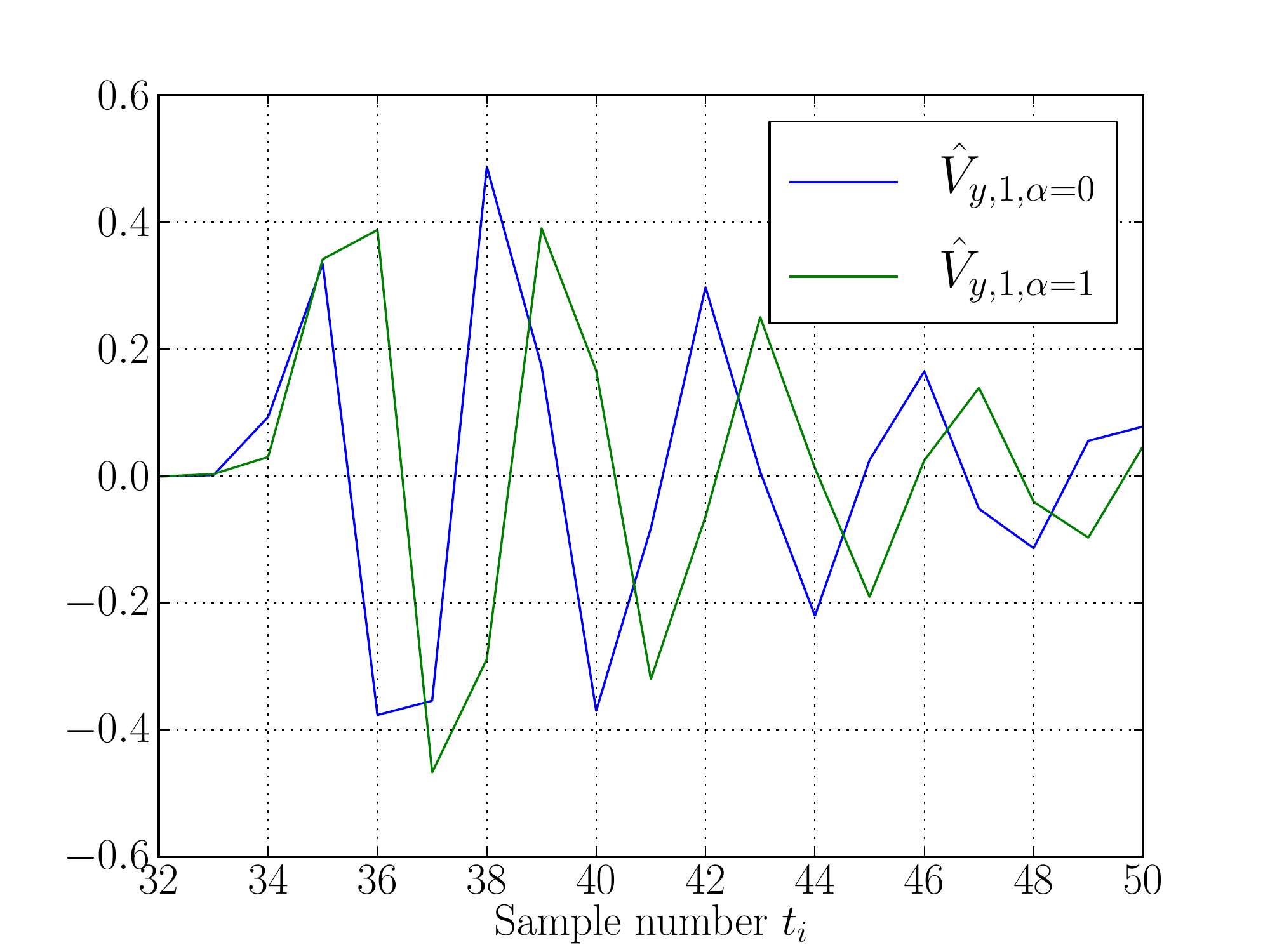} 
\caption{The first two principal components of a CBPM signal plotted together 
.}
\label{fig:figure6}
\end{figure}

Figure~\ref{fig:figure5} shows the result of applying PCA to nonzero IF CBPM digitised data. 
The components are sorted from top to bottom by the explained variance ratio (EVR)~\cite{scikit-learn}, a factor indicating what fraction of the variation in the source data a particular component accounts for. Along with each component the fast Fourier transform of the component FFT(${\bf W}_{j*}^{\bf T}$) is also plotted. 
The first two 
components match the expected shape of the dipole mode signal, and their frequency spectra peak at the expected frequency. The other components contain transients 
and interference signals. The first two components are plotted again in Figure~\ref{fig:figure6}, where it can be clearly seen (apart from the initial transient part) that they are in a 90$^{\circ}$ phase relation. It needs to be noted, that even though the components are orthogonal, they are not necessarily independent, hence, the position  information may still be contained in more than one component, therefore, similar methods, such as Independent Component Analysis (ICA) may be considered for this application.

\begin{figure}[htb]\centering
\includegraphics*[width=120mm]{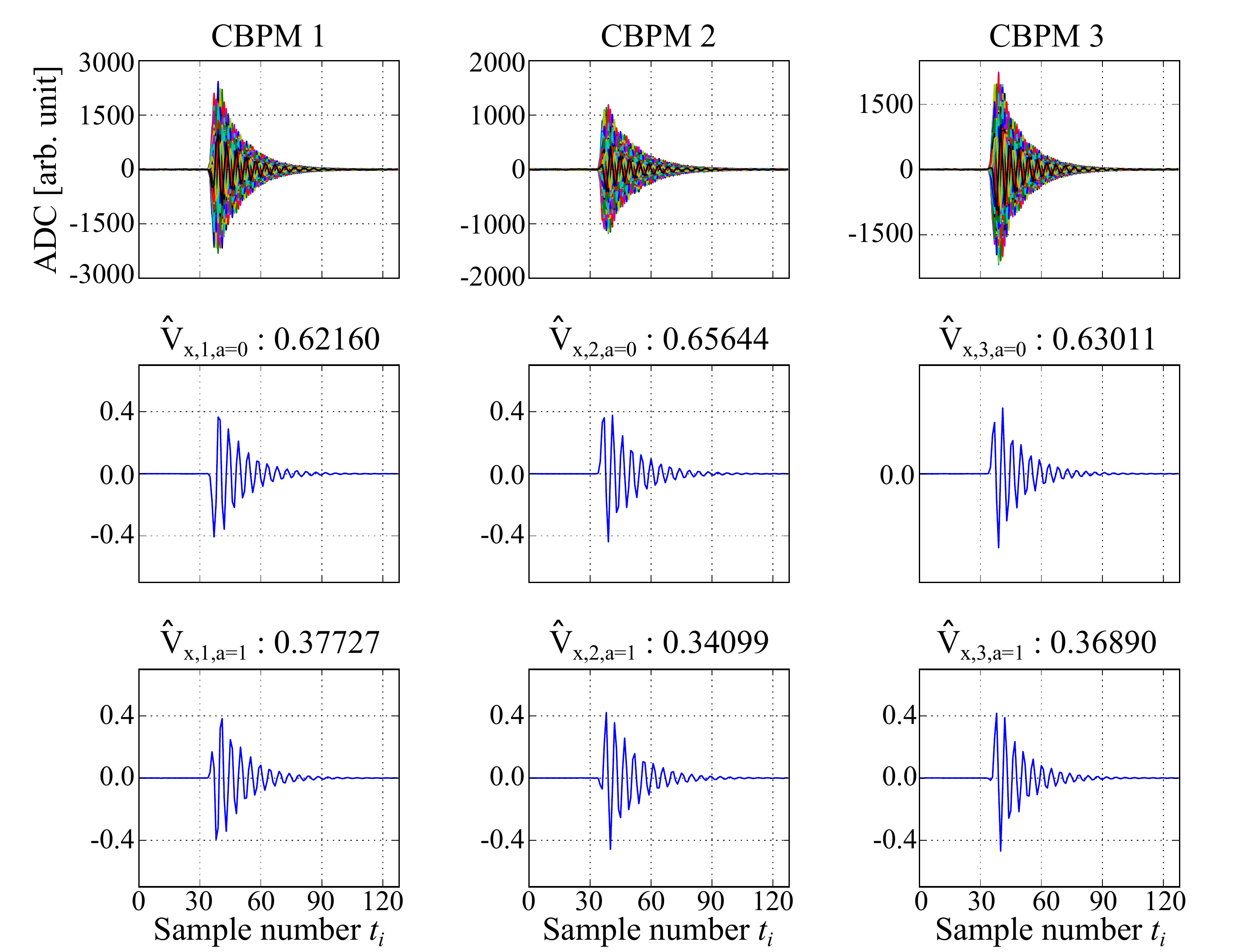}
\caption{PCA components for nonzero IF signals. Raw data (first row), the first component of the signal (second row), the second component
of the signal (last row). } 
\label{fig:figure7}
\end{figure}

The EVR of the first two components depends on the interplay between the beam arrival time and the sampling clock and trigger but the sum of their EVR is close to 100\%. 
This is also true for all of the triplet test CBPMs as shown in Figure~\ref{fig:figure7}. The top row shows raw data for multiple beam passes and the second and the third rows the first and the second principal components respectively for all three CBPMs.
Since the first two components are dipole-like signals and are orthogonal, at least in the body of the waveform, they can be used as a basis for $I\/Q$ demodulation. Their dot products with the measured waveforms provide two values that can be turned into a position reading by applying the rotation and scale previously obtained by calibrating to a known offset, as in conventional processing. 

\begin{figure}[htb]\centering
\includegraphics*[width=120mm]{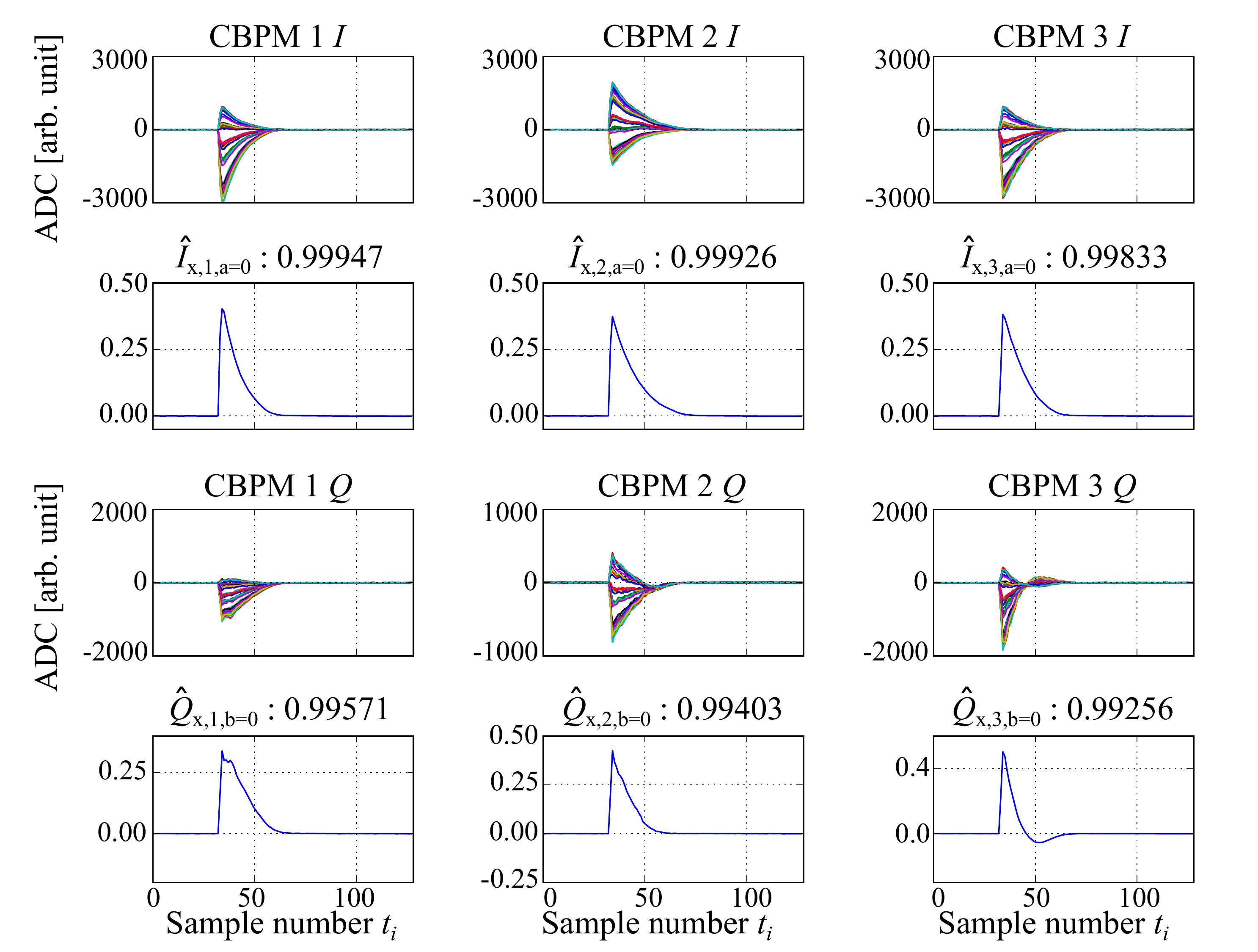}
\caption{PCA components for $I$ and $Q$ signals. Raw $I$ data (first row), the first component of the $I$ signal (second row), raw $Q$ data (third row), the first component of the $Q$ signal (last row).}
\label{fig:figure8}
\end{figure}
Figure~\ref{fig:figure8} shows PCA applied to $I\/Q$ demodulated data. 
In this case only one principal component for each of the $I$ and $Q$ signals is selected, as its EVR is already close to 100\%. The $I\/Q$ basis is already fixed by the LO, but the shape rejection of the unwanted signals still benefits the processing. Taking a dot product with the principal component essentially works as a convolution filter on the waveform, with an important difference that the shape of the filter is derived from the data itself and is known to embrace the useful information -- in this case, the beam position.

\subsection{Calibration}
\begin{figure}[htb]
\centering
\includegraphics*[width=140mm]{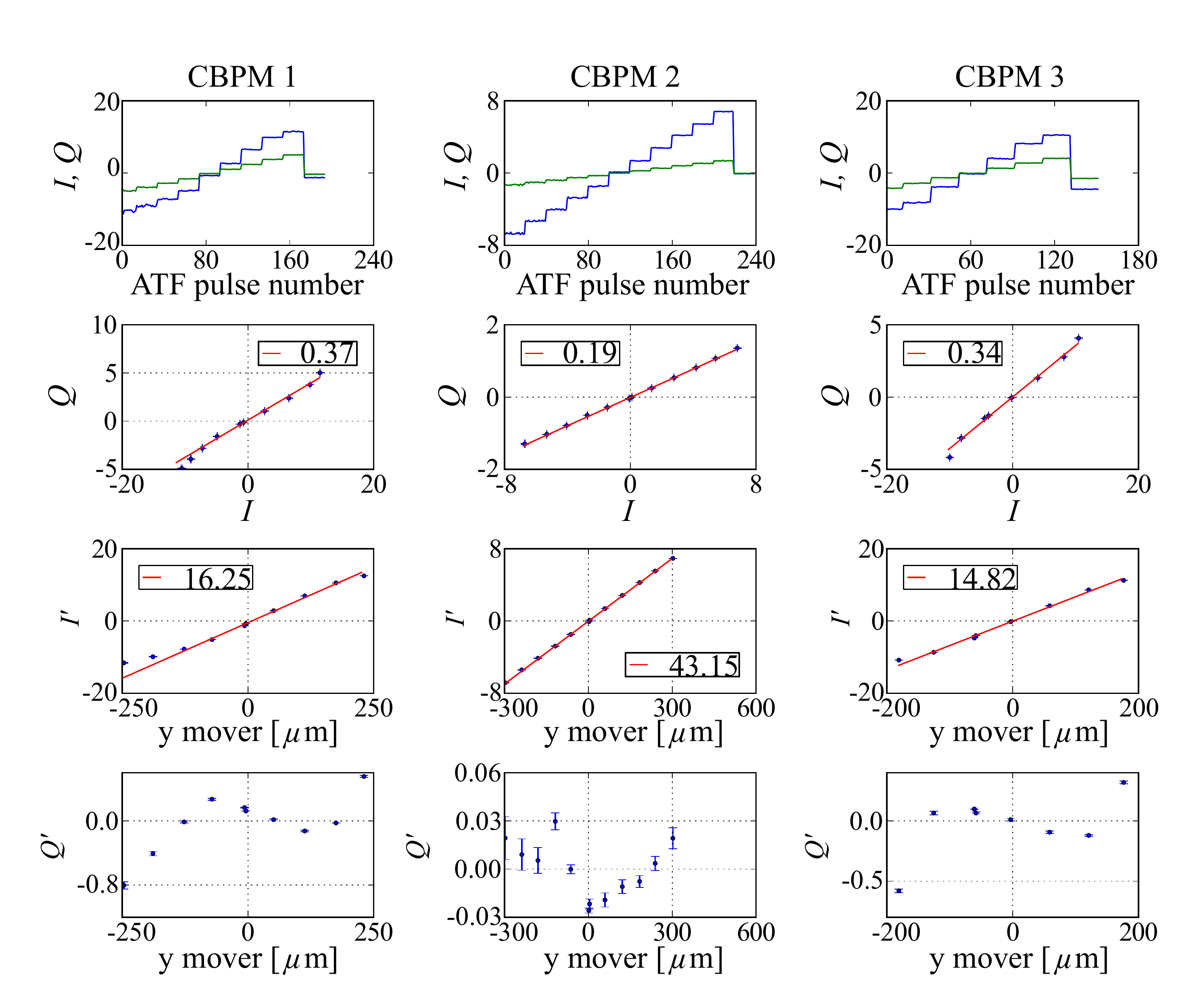}
\caption{A mover calibration for $I\/Q$ data produced using the PCA method, vertical direction, zero-IF demodulation. $I\/Q$ data vs. pulse number (top row), $I\/Q$ data in the $I\/Q$ plane (second row), rotated data vs. mover position (third row), residual rotated $Q$ data (bottom row). The numbers 
in the legends indicate the $\theta_{IQ}$ and $S$ values extracted from a linear fit.}
\label{fig:figure9}
\end{figure}

Figure~\ref{fig:figure9} shows an example calibration for all 3 CBPMs in the setup by applying a known offset using a mover and measuring the rotation $\theta_{IQ}$ of the position data in the $I\/Q$ plane and the position scale $S$. 
The second raw of Figure~\ref{fig:figure9}  shows the $\theta_{IQ}$ measured by fitting $Q$ against $I$ and the third raw shows a similar 
measurement for the position scale $S$. The measured scales indicate a lower sensitivity of CBPM2 compared to the other two cavities. 
This reflects in a higher residual measured for this CBPM in the next section.
The results are very similar to those observed with conventional signal processing, and, in line with Equation~\ref{eq:dipole}, position data lies on a straight line in the $I\/Q$ space.

\section{Performance}\label{sec:results}
\begin{figure}[htb]\centering
\includegraphics*[width=140mm]{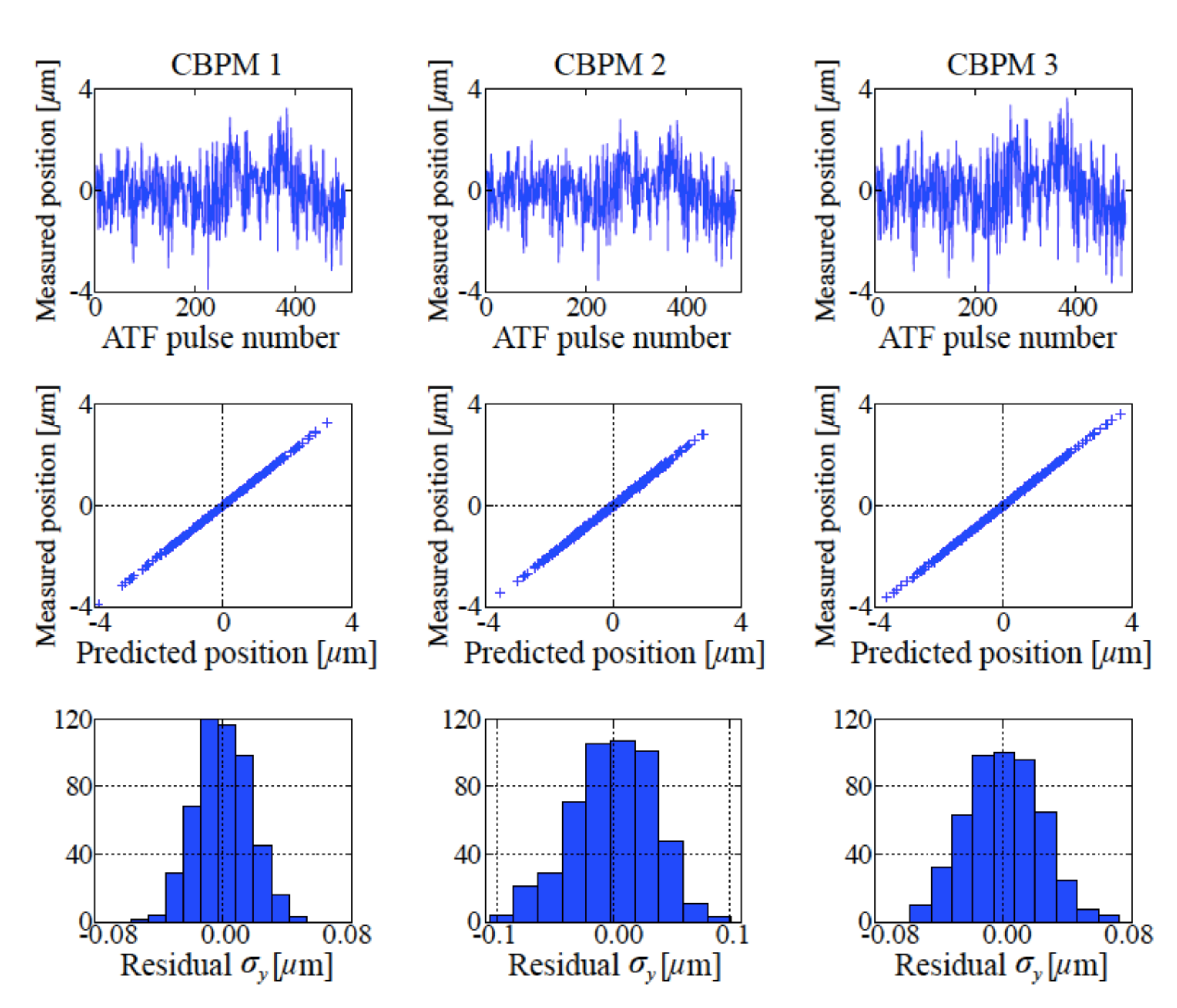}
\caption{Plots for the resolution data from processed using the PCA method.}
\label{fig:figure10}
\end{figure}

The performance of the PCA method was compared to more conventional methods such as 
digital down conversion (DDC) for non-zero IF signals, and simpler single-point readings with and without 
additional filtering and waveform integration in case of analog demodulated $I\/Q$ signals~\cite{kimthesis}. The resolution 
was taken as a basic performance indicator. It was measured as the root-mean-square (RMS) residual 
between the measurement provided by one CBPM in the triplet and the prediction made by the two spectator CBPMs 
as illustrated in Figure~\ref{fig:figure10}. Matrix inversion on the measured data using SVD 
provided correlation coefficients for this prediction.
Figure~\ref{fig:figure10} shows the resolution data from the triplet CBPM system processed with the PCA based method. 
Each column of the figure corresponds to a CBPM in the system. The first row of the plots shows the measured beam position data used for 
the resolution measurement, the second row shows the predicted positions versus the measured ones, and the third row 
shows the residuals between the measured and predicted positions.

Attenuation had to be used during the calibration due to the high sensitivity, and inherently a small offset range of the system compared to the beam jitter at the location for the nominal gain.
Resolution data were taken both with and without 20~dB front-end attenuation (Table~\ref{tab:tab1}). The PCA based processing gives results similar to conventional methods applied to the same data: about 100~nm and 4~nm with and without the attenuation respectively. It should be noted that at maximum sensitivity saturation of the processing electronics may cause a reduction of the position sensitivity at large offsets, resulting in an apparently better resolution measurement, and so the absolute numbers should be treated as indicative. Some more detail on this can be found in~\cite{kimthesis}.

\begin{table}[hbt]
\centering
\caption{Summary of vertical resolution for zero IF signals with and without 20 dB attenuation.}
\begin{tabular}{l|l|l|l|l}
\hline \hline
                                                    &  Single point   & Filter   & Integration & PCA  \\ \hline
with 20 dB attenuation [nm]        & 235                 & 104     & 98             & 98 \\ \hline
without 20 dB attenuation [nm]   & 5.5                  & 3.6      & 3.7             & 4.3  \\
\hline \hline
\end{tabular}
\label{tab:tab1}
\end{table}

\section{Conclusions}\label{sec:conclusion}

A method of processing cavity beam position monitor data based on Principal Component Analysis (PCA) has been successfully tested on the example of a high resolution 3-cavity test system installed in the ATF2 beam line. System performance has been assessed by analysing the residuals between the prediction made by 2 spectator CBPMs and the actual measurement obtained by the third one. A similar performance compared to other more conventional processing methods has been observed.

The most flexible method among the ones tested is the digital down-conversion -- it allows to compensate (to a certain degree) for timing and temperature induced changes in the system. However, it requires 2 additional calibration parameters to determine the beam position compared to the PCA-based processing:
the digital LO frequency and sampling point.
Even for a small test system of 3 cavities + reference that means 16 parameters, $4 ({\rm cavities}) \times 2 ({\rm parameters}) \times 2 ({
\rm directions})$, that require additional beam data and effort. The PCA technique, on the other hand, simplifies the whole process of processing and calibration, which is an important feature in the context of large systems, such as future Linear Colliders. This comes with reduced flexibility and most probably higher sensitivity to long-term effects, such as timing drifts. However, the long term stability for this processing still needs to be understood, and some of the experience with conventional methods may be transferred to this new technique.

There are other methods similar to PCA for constructing a signal basis set, that may be suitable for CBPM processing. These include the independent 
component analysis (ICA) and its variants. Also, a basis set can be generated including the known mover positions 
or beam orbit offsets, so that instead of maximising the variance, a least squares problem is solved.

\acknowledgments
We would like to express our gratitude to all the operators, collaborators, and support staff in the ATF2 group. 
%This work is supported by Science and Technology Facilities Council, U. K., and EuCARD project
%funded by the European Commission within the Framework Program 7, under Grant Agreement No. 227579. 
%We acknowledge the support by the World Class University (Grant No. R32- 20001).

%\bibliographystyle{ieeetr}
%\bibliographystyle{plain}
%\bibliography{pcaPaper}% Produces the bibliography via BibTeX.
\end{document}